\documentclass[letter]{article}
\usepackage{jheppub}
\usepackage{import}
\usepackage{amsmath,amsfonts,relsize}
\usepackage[nameinlink,capitalize]{cleveref} 

\usepackage{graphicx}
\usepackage{dcolumn}
\usepackage[mathlines]{lineno}
\usepackage[caption=false]{subfig}
\usepackage{diagbox}
\usepackage{tabularx}
\usepackage{siunitx}
\usepackage{bm,color,xcolor}
\usepackage{tikz}
\usepackage[compat=1.0.0]{tikz-feynman}

\usepackage{floatrow}
\usepackage{comment}

\usepackage[rightcaption]{sidecap}
\usepackage{fontawesome} 
\sidecaptionvpos{figure}{c}
 
\newfloatcommand{capbtabbox}{table}[][\FBwidth]

\usepackage{enumitem}

\usepackage{xspace}

\usepackage{slashed,physics} 
\usepackage{float}
\usepackage{multirow}
\usepackage{mathtools}
\usepackage{appendix}
\usepackage[colorlinks=true
,urlcolor=blue
,anchorcolor=blue
,citecolor=blue 
,filecolor=blue
,linkcolor=blue
,menucolor=blue
,linktocpage=true
,pdfproducer=medialab
,pdfa=true
]{hyperref}
\usepackage{bm,color,color,soul}
\definecolor{UMNRed}{RGB}{144, 0, 33}
\definecolor{UMNGold}{RGB}{255 204 51}
\usepackage{lineno}

\newcommand{\beq}{\begin{equation}}
\newcommand{\eeq}{\end{equation}}
\newcommand{\bea}{\begin{eqnarray}}
\newcommand{\eea}{\end{eqnarray}}

\usepackage[arrows]{ezedits} 

\defineEditQuick{ZL}{UMNRed}

\title{Z Boson Radiative Decay $Z\rightarrow \mu^+\mu^-\gamma$ at the LHC
}

\author[a]{Yifan Fei,}
\author[b]{Peiran Li,}
\author[b]{Zhen Liu,}
\author[b,c]{Maxim Pospelov}

\affiliation[a]{Department of Physics and Center for Field Theory and Particle Physics,
Fudan University, Shanghai 200438, China}
\affiliation[b]{School of Physics and Astronomy, University of Minnesota, Minneapolis, MN 55455, USA}
\affiliation[c]{William I. Fine Theoretical Physics Institute, School of Physics and Astronomy,
University of Minnesota, Minneapolis, MN 55455, USA}

\emailAdd{24110190018@m.fudan.edu.cn}
\emailAdd{li001800@umn.edu}
\emailAdd{zliuphys@umn.edu}
\emailAdd{pospelov@umn.edu}

\preprint{\small UMN-TH-4525/26,~FTPI-MINN-26-08}

\abstract{
We study the radiative decay of the $Z$ boson, $Z \to \mu^+\mu^-\gamma$, at the LHC, providing both Standard Model (SM) precision analysis and new physics projections. With detailed analysis of Run-2 and future HL-LHC performances, we demonstrate that this decay mode can be measured with a statistical precision at the sub-percentage level. From existing Run-1 data, we extract $\text{Br}^\text{fid}(Z \to \mu\mu\gamma) = (3.34 \pm 0.016)\times 10^{-4}$. We further explore the sensitivity of this channel to axion-like particles (ALPs) and to an anomalous $U(1)_X$ gauge boson coupled to the muon. Both scenarios feature resonant structures in the dimuon invariant mass spectrum within the $Z \to a/X + \gamma \to \mu^+\mu^-\gamma$ final state. Our results show that the radiative $Z$ decay provides a clean and statistically powerful probe of such leptophilic new physics, extending the current collider reach for ALPs and anomalous gauge forces down to $g_X \sim \mathcal{O}(10^{-3})$. This study highlights the potential of rare electroweak gauge boson decays as precision tests of the SM and sensitive probes of new interactions at the LHC. 
}

\begin{document}
\maketitle

\section{Introduction}
The Standard Model (SM) of particle physics has been remarkably successful in describing fundamental interactions, particularly in the electroweak sector, which has been extensively tested within recent decades. Among the electroweak gauge bosons, the $Z$ boson plays a central role and is produced copiously at LEP~\cite{ALEPH:2005ab}
and the LHC~\cite{CMS:2013zfg,CMS:2014pkt,CMS:2024myi,CMS:2025sgv,CMS:2012exm,CMS:2014xja,CMS:2020gtj,CMS:2020hjs,ATLAS:2017bcd,ATLAS:2016bxw,ATLAS:2012bra}, providing unique opportunities to investigate rare decay modes with unprecedented precision, such as $Z\to\gamma\gamma\gamma,\ell\ell\ell\ell$ studies~\cite{ATLAS:2015rsn,ATLAS:2021kog,CMS:2017dzg}. In this work, we investigate the radiative decay $Z \to \mu^+ \mu^- \gamma$, which provides not only a test of the SM predictions but also a potential probe of various new physics scenarios with sub-electroweak scale particles in the spectrum.

The decay $Z \to \mu^+ \mu^- \gamma$ is a SM process that involves final-state radiation (FSR) and is sensitive to detector resolution, phase-space cuts, and electroweak corrections. Although it has been studied at previous $e^+e^-$ colliders such as LEP, only upper limits on the branching ratio have been reported~\cite{OPAL:1991acn} due to the low number of $Z$ boson production. The large 
$Z$ boson dataset collected by the LHC~\cite{CMS:2013zfg,CMS:2014pkt,CMS:2024myi,CMS:2025sgv,CMS:2012exm,CMS:2014xja,CMS:2020gtj,CMS:2020hjs,ATLAS:2017bcd,ATLAS:2016bxw,ATLAS:2012bra,ATLAS:2016oxs} now allows for a precise measurement of this channel.

Furthermore, the $\mu^+ \mu^- \gamma$ final state provides a compelling probe for beyond the Standard Model (BSM) phenomena. In particular, axion-like particles (ALPs) have been widely explored across a broad range of experimental frontiers~\cite{ParticleDataGroup:2024cfk}. Current collider experiments, especially at the LHC, have primarily searched for ALPs through the di-photon resonance channel, probing masses up to the TeV scale~\cite{dEnterria:2021ljz,Mimasu:2014nea,Knapen:2016moh,Jaeckel:2012yz}. However, ALPs may couple not only to electroweak gauge bosons but also to leptons~\cite{Zhitnitsky:1980tq,Dine:1981rt,Bauer:2017ris,Bauer:2021mvw,Bauer:2020jbp,Co:2022bqq}, enabling exotic decay modes such as $Z \to a \gamma \to \mu^+ \mu^- \gamma$. 
In addition to ALPs, another class of BSM models can have an anomalous dark $U(1)_X$ force that couples to muons. Such a gauge interaction is generically anomalous within the Standard Model particle content, requiring new heavy fermions to restore gauge consistency. The loop-induced coupling between the $Z$ boson, the photon, and the new dark gauge boson $X$ gives rise to a radiative decay channel $Z \to X\gamma  \to \mu^+\mu^-\gamma$, leading to the same experimental signature as the ALP case but with a distinct theoretical framework. This makes the radiative $Z$ decay an excellent probe for both pseudo-scalar and vector-type leptophilic extensions of the Standard Model.

In this work, we perform a detailed study of the $Z \to \mu^+ \mu^- \gamma$ process at the LHC, assessing both the SM predictions and potential deviations induced by ALPs or anomalous dark forces. We provide a comprehensive analysis based on Run-2 and projected HL-LHC setup, including parton-level simulations, detector-level cutflows, and sensitivity projections. Additionally, we extract the fiducial branching ratio of the radiative $Z$ decay from a Run-1 study~\cite{CMS:2015vap}, obtaining
$(3.34 \pm 0.016)\times 10^{-4}$. Our results establish the precision potential of measuring the fiducial branching ratio and demonstrate the power of this channel to search for ALPs and muonic dark forces at the current LHC and in future HL-LHC runs.

The paper is organized as follows. In \autoref{sec:fiducial_Br}, we present the SM prediction for the branching ratio of the radiative $Z$ decay and define the fiducial branching ratio as the relevant observable for experimental measurements. The projected sensitivity of the $Z$ decay branching ratio measurement at the LHC is investigated in \autoref{sec:SM_Z_measurement}. Furthermore, this decay channel is utilized to impose constraints on ALPs with several benchmarks in \autoref{sec:ALP} and anomalous dark force in \autoref{sec:dark_U1}. Finally, we provide a summary in \autoref{sec:conclusions}.

\section{Fiducial Branching Ratio}
\label{sec:fiducial_Br}
\begin{figure}[h]
    \centering
    \includegraphics[width=0.495\textwidth]{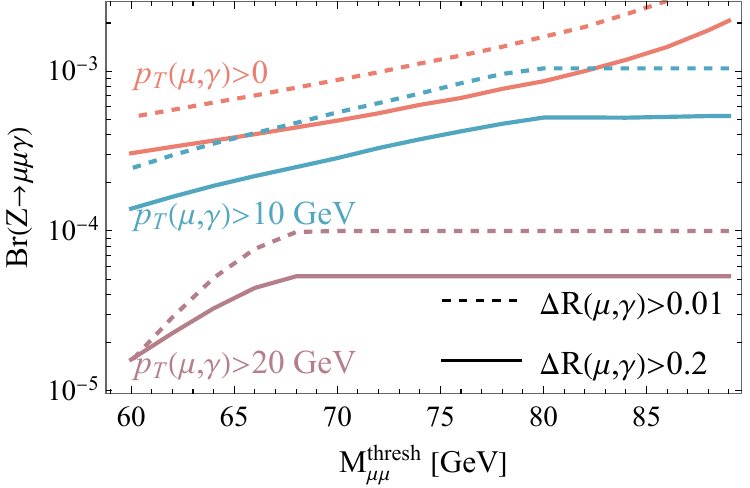}
    \caption{The branching ratio of $Z \to \mu\mu\gamma$ decay channel as a function of dimuon invariant mass $\text{M}^{\rm thresh}_{\mu\mu}$ threshold value with different $\Delta R(\mu,\gamma)$, $p_T$ cuts\,.}
    \label{fig:Br}
\end{figure}

Fixed-order calculation of $Z \to \mu^+\mu^-\gamma$ contains large infrared logarithms, which require resummation with higher-order corrections. In this work, for simplicity and clarity, we focus on the fiducial branching ratio, for which a fixed-order calculation with appropriate (sufficiently strong) phase-space cuts is sufficient.   
The $Z\to\mu\mu\gamma$ branching ratio as a function of dimuon invariant mass threshold $\text{M}_{\mu\mu}^\text{thresh}$ is plotted in \autoref{fig:Br} (computed via {\tt MadGraph5\_aMC@NLO}~\cite{Alwall:2014hca}), where $\text{M}_{\mu\mu}^{\rm thresh}$ denotes the upper cut imposed on the dimuon invariant mass. 
\autoref{fig:Br} illustrates how the fiducial rate is shaped by the soft/collinear phase space under different $p_T$ and $\Delta R$ selections.
With a hard-photon requirement, the dimuon mass has a kinematic endpoint in the $Z$ rest frame,
$\text{M}_{\mu\mu}^2 \simeq m_Z^2-2m_ZE_\gamma$; hence $p_T(\gamma)$ cuts imply $E_\gamma\gtrsim p_T^{\rm cut}$ and thus $\text{M}_{\mu\mu}\le \text{M}_{\mu\mu}^{\max}$. 
As a result, the integrated branching ratio becomes flat on the right once $\text{M}_{\mu\mu}^{\rm thresh} \gtrsim \text{M}_{\mu\mu}^{\max}$, and it decreases when lowering $\text{M}_{\mu\mu}^{\rm thresh}$ below this endpoint.
The dependence on $\Delta R(\mu,\gamma)$ reflects the collinear enhancement of final-state radiation: relaxing $\Delta R$ increases the rate.
In this work we adopt $\Delta R(\mu,\gamma)>0.2$ as our standard benchmark (together with a conservative $p_T(\gamma)$ cut) to stay in the fixed-order-safe regime.\footnote{In the analyses of \cref{sec:SM_Z_measurement}, we do not impose an additional cut on $\text{M}_{\mu\mu}$, since our $p_T(\mu,\gamma)$ and $\Delta R(\mu,\gamma)$
requirements already remove the soft and collinear regions and are sufficient for our fixed-order analysis.}
It indicates that imposing a threshold of $p_T(\mu,\gamma) \gtrsim 10~\text{GeV}$ and $\Delta R(\mu,\gamma)>0.2$ results in a branching ratio of approximately $\sim 10^{-4}$.
Given the huge number of $Z$ bosons produced at the LHC ($\sim 10^{10}$ in Run-2), such a channel can be clearly measured.

Similar to \cite{Fei:2024qtu}, the fiducial branching ratio, which can be experimentally measured, is defined as
\begin{align}
    \text{Br}^\text{fid}=\frac{\sigma^\text{fid}(pp\to Z\to\mu\mu\gamma)}{\sigma(pp\to Z)}=\frac{\epsilon^\text{cuts} \times\sigma(pp\to Z\to\mu\mu\gamma)}{\sigma(pp\to Z)}\, , \label{eq:Br_fid}
\end{align}
where $\epsilon^\text{cuts}$ corresponds to the detector-level selection efficiency, and $\sigma(pp\to Z\to\mu\mu\gamma)$ is the signal cross section (after parton-level cuts and jet-merging, which will be discussed in \autoref{sec:parton_lv}). Throughout this work, when referring to the signal process $pp\to Z\to\mu^+\mu^-\gamma$, we use the narrow-width approximation as an operational definition of the resonant on-shell $Z$ contribution, 
$\sigma(pp\to Z\to\mu\mu\gamma)=\sigma(pp\to Z)\,{\rm Br}(Z\to\mu^+\mu^-\gamma)$.
In \autoref{eq:Br_fid}, $\sigma^{\rm fid}(pp\to Z\to\mu\mu\gamma)$ denotes the resonant on-shell $Z$ contribution evaluated in the narrow-width approximation. The experimentally measured fiducial rate is instead the inclusive $\mu\mu\gamma$ rate, which also contains the non-resonant Drell--Yan continuum. The extraction of the resonant contribution therefore requires a theory input
\[
f_Z
\equiv
\frac{
\sigma^{\rm fid}(pp\to Z\to\mu^+\mu^-\gamma)
}{
\sigma^{\rm fid}(pp\to\mu^+\mu^-\gamma)
},
\]
so that
\[
{\rm Br}^{\rm fid}(Z\to\mu^+\mu^-\gamma)
=
\frac{
\sigma^{\rm fid}(pp\to\mu^+\mu^-\gamma)
}{
\sigma(pp\to Z)
}f_Z =\frac{\sigma^\text{fid}(pp\to Z\to\mu\mu\gamma)}{\sigma(pp\to Z)}.
\]
At leading order, this is equivalent to estimating the non-resonant continuum as the difference between the inclusive $\mu^+\mu^-\gamma$ sample and the narrow-width resonant sample. In principle, the inclusive $pp\to \mu\mu\gamma$ rate also contains the interference between the resonant on-shell $Z$ amplitude and the non-resonant Drell–Yan amplitudes. Under our operational separation, this interference is absorbed into the background component defined by the difference between the inclusive sample and the narrow-width resonant contribution. The on-shell regime interference effects are small in the $Z$-pole region, where the resonant contribution dominates, and the interference largely cancels upon integration across the resonance upon analysis of the relative phases and phase-space overlap of different contributing amplitudes.

\section{Standard Model $Z$ Boson decay precision measurement}
\label{sec:SM_Z_measurement}
The radiative decay branching ratio of the $Z$ boson, $\text{Br}(Z\to \mu\mu\gamma)$, has been searched at electron-positron colliders~\cite{OPAL:1991acn}, where only an upper limit on the fiducial branching ratio has been established ($<5.6\times 10^{-4}$). The huge amount of $Z$ bosons produced at the LHC should significantly improve the measurement. In this section, we study the projected sensitivity of $Z\to \mu\mu \gamma$ branching ratio for both Run-2 and HL-LHC configurations. Run-2 benchmark is treated as 13~TeV center-of-mass energy with $140~\text{fb}^{-1}$ integrated luminosity following ATLAS/CMS detector performance~\cite{CMS:2021yvr,ATLAS:2020gty}; HL-LHC benchmark is treated as 14~TeV energy with $3~\text{ab}^{-1}$ integrated luminosity following future HL-LHC performance~\cite{Apollinari:2015bam}. Additionally, we directly extract the $Z$ boson radiative decay branching ratio via reinterpretation of a Run-1 study~\cite{CMS:2015vap}.

\subsection{Signal and Background Consideration}
\label{sec:parton_lv}
The leading channel for on-shell $Z$ boson production $u\bar{u}/d\bar{d}\to Z$ is convoluted under the proton parton distribution function (PDF).  For the signal process, we require the $Z$ boson decays into $\mu\mu\gamma$ with enough photon energy ($p_T(\gamma)>5~\text{GeV}$ and $\Delta R(\mu\gamma)>0.2$) to avoid the soft collinear divergence, which should be considered within resummation of $Z$ boson di-lepton decay. 
Focusing on the fixed-order perturbative regime, the parton-level cut for the Run-2 (HL-LHC) benchmark is
\begin{align*}
    &p_T(\mu)>5~\text{GeV},~p_T(\gamma)>5~\text{GeV},~p_T(j)>20~\text{GeV},~|\eta(\mu)|<2.5(2.8),~|\eta(\gamma)|<2.5,~|\eta(j)|<5,\\
    &\Delta R(\mu\mu,\mu\gamma,j\gamma)>0.2\, 
\end{align*}
where $\Delta R=\sqrt{(\Delta\phi)^2+(\Delta\eta)^2}$. The signal events then pass through the interfaced {\tt Pythia8}~\cite{Bierlich:2022pfr} for parton shower and jet matching. We match up to one jet and choose the merging scale to be $15~$GeV. After jet matching, the truth-level cross section\footnote{Truth-level cross section is the cross section after both parton-level cuts and one-jet merging process.} of the signal process is shown in \autoref{Table:cutflow_table_Run2} (\autoref{Table:cutflow_table_HLLHC}).

The primary backgrounds can be summarized as follows:
\begin{itemize}
    \item $pp\to \mu\mu\gamma$,
    \item $pp\to W^+W^-\gamma\to \mu\nu_\mu \mu \bar{\nu}_\nu \gamma$,
    \item $pp\to Z\mu\mu \gamma\to \nu\bar{\nu} \mu\mu  \gamma$,
    \item $pp\to W\mu\mu\gamma\to \mu\nu_\mu \mu\mu \gamma$.
\end{itemize}

The first process, $pp\to\mu\mu\gamma$, includes both signal and background processes. In our case, the main background process is $q\bar{q}\to \gamma^*/Z^{(*)}+ \gamma\to\mu\mu\gamma$, which contains both the Drell-Yan process with a hard photon radiation and diphoton production $\gamma^*\gamma$ where one of the photons is slightly offshell to produce dimuon. 

The other three processes are primarily diboson productions ($WW,ZZ,WZ$) in association with a hard photon.  
The most relevant phase-space from these diboson processes involves off-shell massive bosons. For example, we simulate $pp\to Z\mu\mu\gamma$ with $Z\to \nu\bar{\nu}$. 
One could alternatively simulate the full final state directly, such as $pp\to \mu\mu\nu\nu\gamma$, but the result would be nearly identical, since our approach already includes most of the relevant off-shell phase space, which is dominated as one of the $W/Z$ is on-shell. Notice that the last background has three muons in the final state, but one of the muons can be missed at the detector level.

The cross sections of all backgrounds are shown in \autoref{Table:cutflow_table_Run2} and \autoref{Table:cutflow_table_HLLHC}. As higher-order processes, the diboson-production backgrounds are much smaller than the Drell-Yan process as expected, not least because of a smaller PDF as they require higher $\sqrt{\hat{s}}$.

\subsection{Cutflow Analysis at Detector Level}
\label{sec:cutflow_reconstr}
After passing through {\tt Pythia8} for parton shower and jet merging, the truth-level sample is sent to {\tt Delphes3}~\cite{deFavereau:2013fsa,Mertens:2015kba} for a fast detector simulation to include various detector effects. For the Run-2 (HL-LHC) benchmark, we implement the following detector-level cuts before optimizing the kinematics selection: 
\begin{align*}
    &p_T^1(\mu)>20(10)~\text{GeV},~p_T^2(\mu)>10(5)~\text{GeV},~p_T(\gamma)>20~\text{GeV},~|\eta(\mu)|<2.8(2.5),\\ 
    &\Delta R(\mu\mu)>0.2,~
    n(\mu)=2,~n(\gamma)=1,~n(j) = 0
\end{align*}
where $\text{M}_{\mu\mu\gamma}$ is the invariant mass of $\mu\mu\gamma$. We require two muons passing the $p_T$ criteria, one photon with $p_T(\gamma) > 20~ \text{GeV}$ and no jet has $p_T(j)>20~\text{GeV}$. The above selections are based on both the current trigger system from ATLAS and CMS~\cite{CMS:2021yvr,ATLAS:2020gty,CMS:2024aqx,ATL-DAQ-PUB-2019-001, ATLAS:2020gty} and future HL-LHC detector performance~\cite{Ryd:2020ear,Guiducci:2018ogp}

\subsubsection{Run-2 benchmark}
\begin{figure}[ht]
    \centering
    \includegraphics[width=0.496\textwidth]{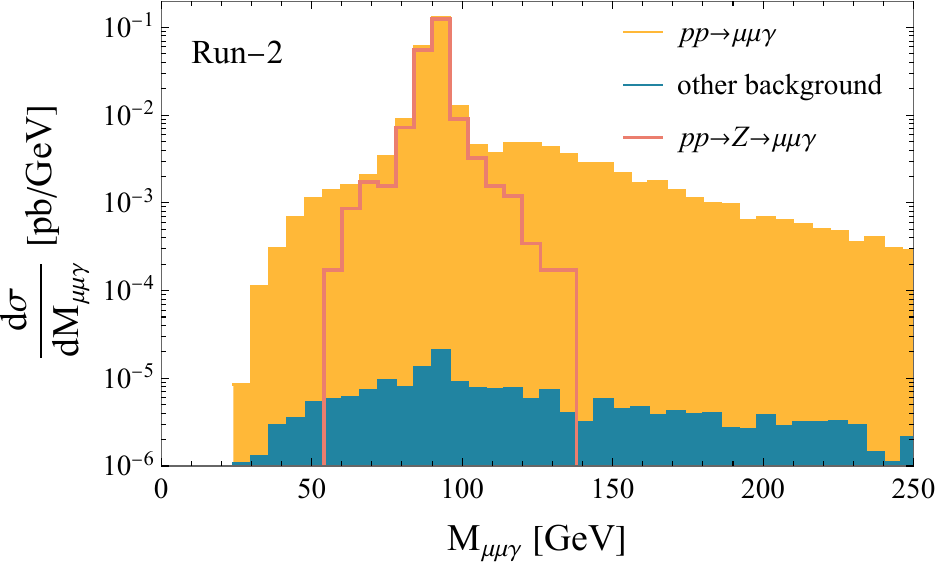}
    \includegraphics[width=0.496\textwidth]{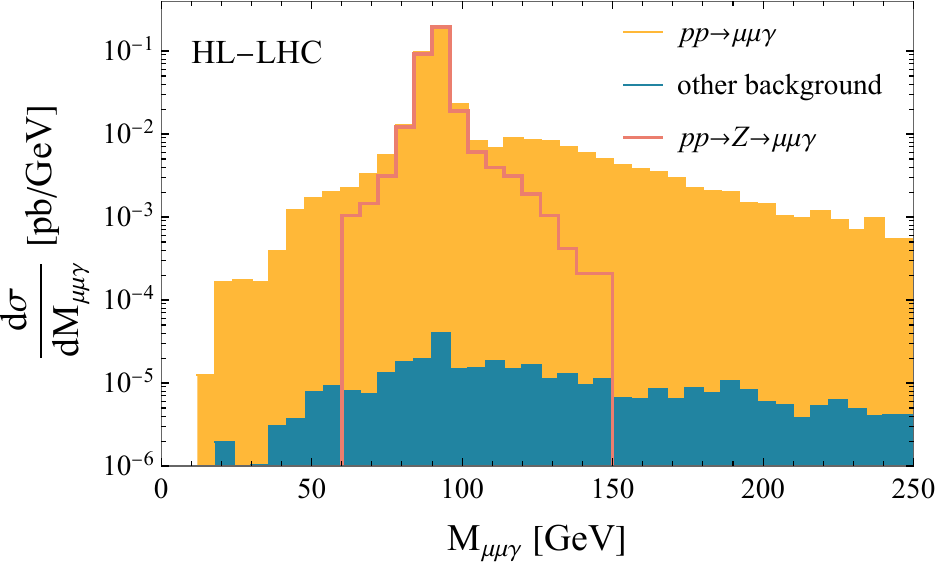}
    \caption{Detector-level distribution of $\text{M}_{\mu\mu\gamma}$ after detector-level cuts (except the mass window selection). The only differences between Run-2 and HL-LHC benchmark are the muon $p_T$ threshold and the muon rapidity coverage at detector-level. The signal process is plotted as a line, because $pp\to\mu\mu\gamma$ already includes the signal contribution.
    }
    \label{fig:mumua_distribution}
\end{figure}

Run-2 benchmark is treated as $\sqrt{s}=13~\text{TeV}$ with $140~\text{fb}^{-1}$ integrated luminosity. Implementing the above detector-level cuts to both signal and background processes, the corresponding cutflow table is shown in \autoref{Table:cutflow_table_Run2}. The cut efficiency in the third column is mainly affected by the difference between the parton-level cuts used during sample generation and the detector-level $p_T$ thresholds. Our sample was generated with $p_T(\mu, \gamma)>5~\text{GeV}$, and the detector-level selection requires $p_T^{1}(\mu) > 20~\text{GeV}$, $p_T^{2}(\mu) > 10~\text{GeV}$, and $p_T(\gamma) > 20~\text{GeV}$. This leads to a low selection efficiency even for the signal process. Moreover, we show the invariant mass of $\mu\mu\gamma$ in \autoref{fig:mumua_distribution} and the $pp\to\mu\mu\gamma$ process dominates the whole spectrum. Because ($pp\to\mu\mu\gamma$) includes both the on-shell $Z$ boson radiative decay and other contributions from other diagrams, we plot the signal process as a red line over the region in yellow. As expected, the red line is highly consistent with the yellow region around the $Z$ pole, which implies that the on-shell $Z$ boson dominates around the $Z$ pole region.

We apply a mass window cut $80<\text{M}_{\mu\mu\gamma}<100~\text{GeV}$ as the kinematics optimization selection and compute the statistical precision $\frac{\sqrt{S+B}}{S}$ where $S$ and $B$ are the signal and backgrounds events after all selections. Notice that after the mass window selection (shown in the last column of \autoref{Table:cutflow_table_Run2}), the signal cross section almost saturates the $pp\to\mu\mu\gamma$ process, which is expected from the $\text{M}_{\mu\mu\gamma}$ distribution. The statistical precision of the fiducial cross section will propagate to the precision of the fiducial branching ratio:
\begin{align}
    \frac{\delta\text{Br}^{\text{fid}}(Z\to\mu\mu\gamma)}{\text{Br}^{\text{fid}}(Z\to\mu\mu\gamma)}= 0.26\% \,.
\end{align}  
The fiducial branching ratio in this analysis is roughly $\sim10^{-5}$ and directly depends on the detector-level cuts, which reflect the detector performance. Therefore, we only report the relative branching ratio precision (only based on statistical uncertainty) for both the Run-2 and HL-LHC analyses. 

\begin{table}[ht]
\centering
\resizebox{1.0\textwidth}{!}{%
\begin{tabular}{|c|c|c|c|c|}
    \hline
     Cross section $[\text{pb}]$ & Truth-level & $n(\mu)=2, n(\gamma)=1$ & $n(j) = 0$ & $80<\text{M}_{\mu\mu\gamma}<100~\text{GeV}$ \\
    \hline
    $pp \to Z \to \mu\mu\gamma$ & $20.5(100\%)$ & $1.91(9.3\%)$ & $1.25(6.1\%)$ & $1.17(5.7\%)$\\
    \hline
    $pp \to \mu\mu\gamma $ & $\num{27.0}(100\%)$ & $\num{2.55}(9.5\%)$ & $\num{1.63}(6.0\%)$ & $\num{1.24}(4.6\%)$\\
    \hline
    $pp \to WW\gamma$ & $0.0216(100\%)$ & $\num{2.46e-3}(11\%)$ & $\num{9.68e-4}(4.5\%)$ & $\num{1.22e-4}(0.56\%)$\\
    \hline
    $pp \to Z\mu\mu\gamma $ & $0.00144(100\%)$ & $\num{7.25e-4}(9.7\%)$ & $\num{9.56e-5}(2.4\%)$ & $\num{6.06e-5}(4.2\%)$\\
    \hline
    $pp \to W\mu\mu\gamma$ & $0.00749(100\%)$ & $\num{7.25e-4}(9.7\%)$ & $\num{1.82e-4}(2.4\%)$ & $\num{7.65e-5}(1.0\%)$\\
    \hline 
    \end{tabular}
}
\caption{Cutflow table of signal and backgrounds under 13 TeV Run-2 benchmark. The $p_T$, $\eta$ and $\Delta R$ cuts are absorbed into $n(\mu)=2, n(\gamma)=1$. Notice that $pp\to\mu\mu\gamma$ includes our signal $pp\to Z\to \mu\mu\gamma$.}
\label{Table:cutflow_table_Run2}
\end{table}

\subsubsection{Run-1 result}
The Run-2 analysis points out that this is an extremely clear channel. We also notice that ATLAS and CMS were using this channel as a calibration on photon identification~\cite{ATLAS:2019qmc,CMS:2020uim}. It is still recommended to refresh the measurement of $Z$ boson radiative decay since the previous measurement via LEP~\cite{OPAL:1991acn} only established upper limits on the fiducial branching ratio. For instance, photon calibration can also be performed independently using the $Z\to e^+ e^- \gamma$ channel.

Fortunately, one of the Run-1 7~TeV studies from CMS~\cite{CMS:2015vap} discussed this $\mu\mu\gamma$ channel. Their $\mu\mu\gamma$ invariant mass distribution includes the total signal events. We extract the signal events within $80<\text{M}_{\mu\mu\gamma}<100~\text{GeV}$ and translate into the $Z$ boson radiative branching ratio
 \begin{align}
    \text{Br}^\text{fid}(Z\to \mu\mu\gamma)=(3.34\pm0.016)\times 10^{-4}
    \label{eq:branching-ratio}
\end{align}
where the uncertainty only comes from statistical uncertainty of signal events $\sqrt{S}$. 
We also note that the ATLAS conference note~\cite{ATLAS-CONF-2022-046}, based on 8 TeV data, reported a $Z\to \ell\ell\gamma$ analysis. Translating its result into a fiducial branching ratio yields $\text{Br}^\text{fid}\sim \mathcal{O}(10^{-4})$, in agreement with above.

\subsubsection{HL-LHC benchmark}
Similar to Run-2 benchmark, HL-LHC benchmark is treated as $\sqrt{s}=14~\text{TeV}$ with $3~\text{ab}^{-1}$ integrated luminosity. Unlike the Run-2 benchmark, where we applied conservative $p_T$ thresholds, the current lepton trigger systems~\cite{CMS:2021yvr,ATLAS:2020gty} allow a lower transverse momentum threshold of $4~\text{GeV}$. We assume this low threshold can be maintained at the HL-LHC ($p_T^1(\mu)>10~\text{GeV},~p_T^2(\mu)>5~\text{GeV}$). The rest of the analysis is similar to the Run-2 benchmark. The cutflow table is shown in \autoref{Table:cutflow_table_HLLHC}, noticing the truth-level cross section is slightly different from the Run-2 analysis simply because of the difference in $\sqrt{s}$ and $|\eta(\mu)|$ at parton-level generation. The kinematics distribution is also plotted in the right panel of \autoref{fig:mumua_distribution}, which has the same behavior as Run-2 benchmark. After the cutflow analysis, we end with a precision of 
\begin{align}
    \frac{\delta\text{Br}^{\text{fid}}(Z\to\mu\mu\gamma)}{\text{Br}^{\text{fid}}(Z\to\mu\mu\gamma)}= 0.043\% \,,
\end{align}  
based on statistical uncertainty.

\begin{table}[ht]
\centering
\resizebox{1.0\textwidth}{!}{%
\begin{tabular}{|c|c|c|c|c|}
\hline
 Cross section $[\text{pb}]$ & Truth-level & $n(\mu)=2, n(\gamma)=1$ & $n(j) = 0$ & $80<\text{M}_{\mu\mu\gamma}<100~\text{GeV}$ \\
\hline
$pp \to Z \to \mu\mu\gamma$ & $24.8(100\%)$ & $2.70(11\%)$ & $2.05(8.3\%)$ & $1.87(7.5\%)$\\
\hline
$pp \to \mu\mu\gamma $ & $\num{31.8}(100\%)$ & $\num{3.51}(11\%)$ & $\num{2.60}(8.2\%)$ & $\num{1.89}(5.9\%)$\\
\hline
$pp \to WW\gamma$ & $0.0258(100\%)$ & $\num{3.69e-3}(14\%)$ & $\num{1.79e-3}(6.9\%)$ & $\num{2.01e-4}(0.78\%)$\\
\hline
$pp \to Z\mu\mu\gamma$ & $0.00173(100\%)$ & $\num{2.11e-4}(12\%)$ & $\num{1.37e-4}(7.9\%)$ & $\num{7.46e-5}(4.3\%)$\\
\hline
$pp \to W\mu\mu\gamma$ & $0.00943(100\%)$ & $\num{1.08e-3}(11\%)$ & $\num{3.91e-4}(4.1\%)$ & $\num{1.54e-4}(1.6\%)$\\
\hline 
\end{tabular}
}
\caption{Cutflow table of signal and backgrounds under 14 TeV HL-LHC benchmark. The $p_T$, $\eta$ and $\Delta R$ cuts are absorbed into $n(\mu)=2, n(\gamma)=1$. Notice that $pp\to\mu\mu\gamma$ includes our signal $pp\to Z\to \mu\mu\gamma$.}
\label{Table:cutflow_table_HLLHC}
\end{table}

\subsubsection{Systematic uncertainty}
The systematic uncertainty at the LHC can be percent level. \cite{CMS:2015vap} reports systematic uncertainties for $\dd\sigma/\dd E_T$ and $\dd\sigma/\dd \Delta R$ that are typically at a few percent level. We therefore expect the systematic uncertainty on the extracted branching ratio to be around percent level or below, given that our observable is a more inclusive rate, in comparison with the finely-binned differential observables.
A full extraction of the fiducial branching ratio at such level of precision would require a dedicated treatment of experimental systematics and theoretical uncertainties, including higher-order electroweak/QED corrections.

We give similar comments on the systematic uncertainties as our previous $W$ boson exotic decay study mentioned~\cite{Fei:2024qtu}. A dominant source comes from uncertainties in the $Z$ boson production rate, arising from parton distribution functions (PDFs), higher-order corrections and luminosity. In both the Run-2 and HL-LHC benchmark studies, the statistical uncertainty is as low as $\sim \mathcal{O}(0.1\%)$, making it essential to minimize the larger systematic uncertainties of $\mathcal{O}(1\%)$. For example, the proton PDF uncertainty can reach approximately 3\%. To mitigate this, one can consider using a ratio of branching fractions as the observable:
\begin{align*}
    \frac{\text{Br}(Z \to \mu\mu\gamma)}{\text{Br}(Z \to \mu\mu)},
\end{align*}
which is relatively insensitive to PDF uncertainty. The uncertainty associated with the muon tagging efficiency, which is one of the main sources of systematic uncertainty in~\cite{CMS:2015vap}, is also expected to be mitigated. 
On the other hand, the photon reconstruction efficiency does not cancel in this ratio and therefore remains one of the leading experimental systematic uncertainties.

\section{Axion-like Particle}
\label{sec:ALP}
\begin{figure}[ht]
    \centering
    \includegraphics[width=0.26\textwidth]{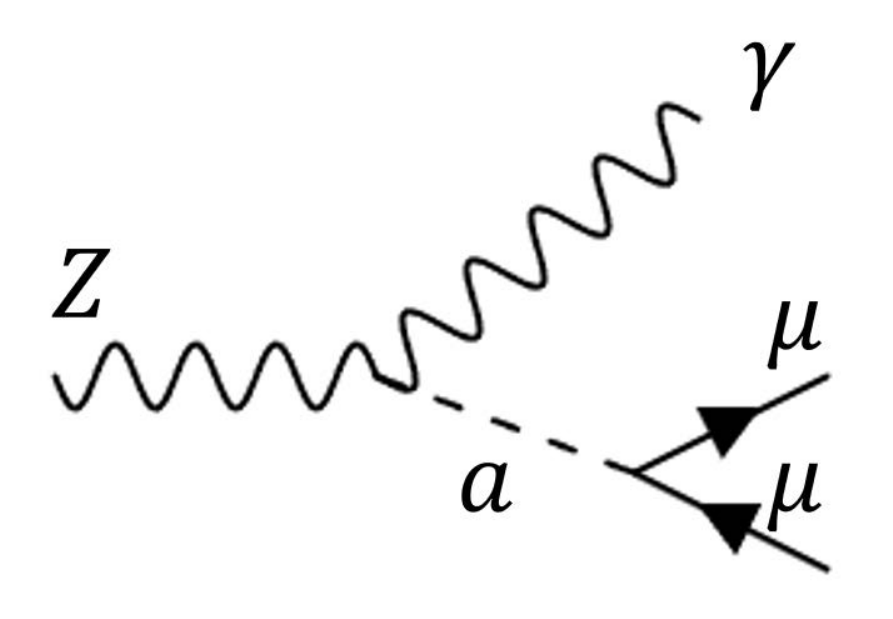}~~~~~~~
    \includegraphics[width=0.26\textwidth]{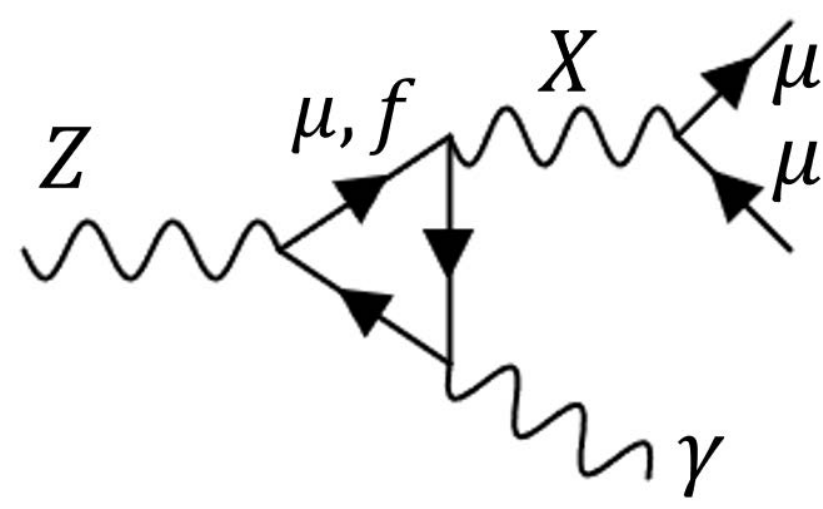}
    \caption{The Feynman diagram of $Z\to \mu\mu\gamma$ from ALP (left) and anomaly-dark-vector benchmark (right).}
    \label{fig:feynman_diagram}
\end{figure}

Axion-like particles (ALPs) are hypothetical pseudoscalar bosons that appear in many extensions of the Standard Model, typically as pseudo-Goldstone bosons of spontaneously broken global symmetries. Unlike the QCD axion, mainly coupled to the QCD sector and motivated by the strong CP problem, ALPs generically may couple to other sectors and have been widely explored across a broad range of experimental frontiers~\cite{ParticleDataGroup:2024cfk}. 
At colliders, and in particular at the LHC, ALP searches have largely focused on di-photon resonances, extending sensitivity to masses at the TeV scale~\cite{dEnterria:2021ljz,Mimasu:2014nea,Knapen:2016moh,Jaeckel:2012yz}. In more general setups, ALPs can interact not only with electroweak gauge fields but also directly with leptons~\cite{Zhitnitsky:1980tq,Dine:1981rt,Bauer:2017ris,Bauer:2021mvw,Bauer:2020jbp,Co:2022bqq}, which can be searched based on $Z\to \mu^+ \mu^- \gamma$. 
Within a kinematic-allowed regime, an on-shell $Z$ boson is able to decay into $a\gamma$, and the ALP further decays into dimuon (shown in the left panel of \autoref{fig:feynman_diagram}):
\begin{align}
    \mathcal{L}\supset c_1\frac{\alpha_1}{8\pi f_a}aB\Tilde{B}+c_2\frac{\alpha_2}{8\pi f_a}aW\Tilde{W}+c_\ell\frac{\partial_\alpha a}{2f_a}\bar{\ell}\gamma^\alpha \gamma_5 \ell \, ,
\end{align}
where $\ell$ stands for $e,\mu$ and $c_1,c_2,c_\ell$ are generic $\mathcal{O}(1)$ Wilson coefficients depending on the UV model. One can also turn on the $\tau$ sector coupling, but the ALP decay branching ratio will be dominated by $\tau \bar{\tau}$ and will be hard to tag in the LHC. 

The decay rates of ALP with $5~\text{GeV}<m_a<90~\text{GeV}$ are
\begin{align}
    &\Gamma(a\to \gamma\gamma)=\frac{1}{2}\times\frac{\alpha_\gamma^2 (c_1+c_2)^2}{128\pi^3 f_a^2} \cdot m_a^3~,\\
    &\Gamma(a \to \mu^+\mu^-) = \frac{c_\mu^2 }{8\pi f_a^2}  \cdot m_\mu^2 m_a\sqrt{1 - \frac{4m_\mu^2}{m_a^2}}\, .
\end{align}
The branching ratio of ALP to electron pairs is negligible, due to the tiny mass of the electron. 

Therefore, we concentrate on the $\mu\mu\gamma$ final state. As indicated in \autoref{Table:cutflow_table_Run2} and \autoref{Table:cutflow_table_HLLHC}, after applying the selection $80 < \text{M}_{\mu\mu\gamma} < 100~\text{GeV}$, the dominant SM background arises from the radiative decay of the $Z$ boson, which was discussed in the preceding section. It now stands as the primary background.

\subsection{Simulation and Projections}
\begin{figure}[ht]
    \centering
    \includegraphics[width=0.495\textwidth]{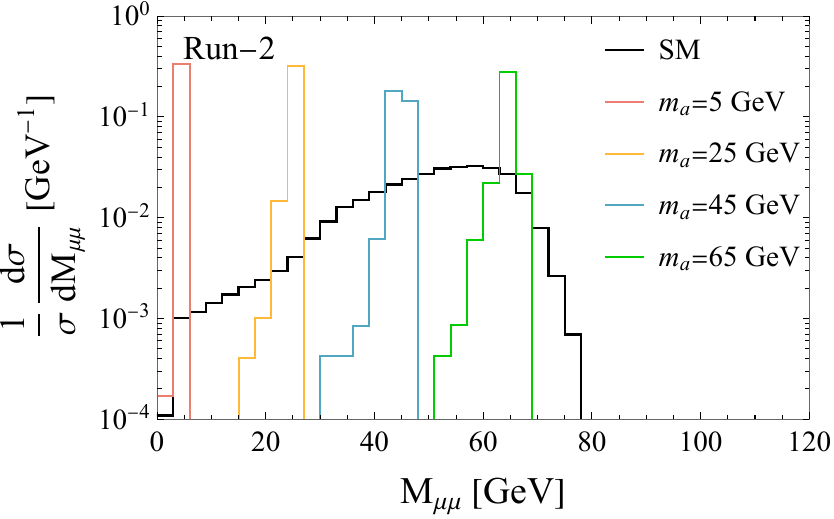}~
    \includegraphics[width=0.495\textwidth]{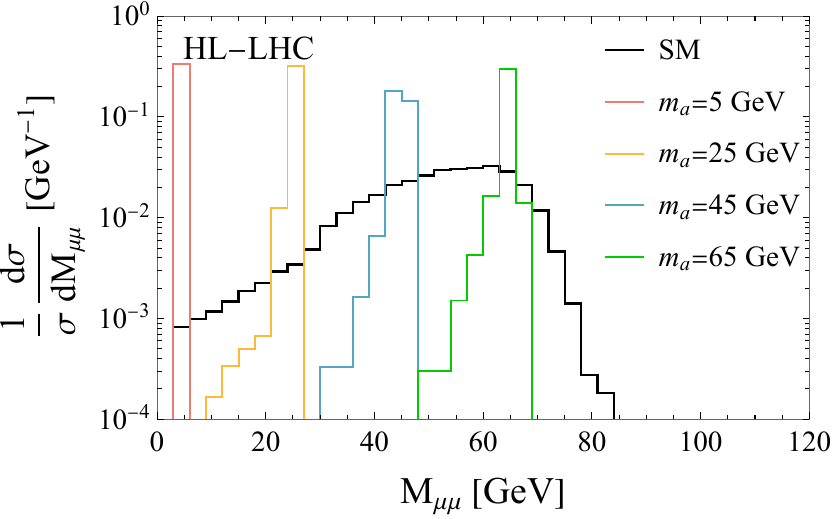}
    \caption{Detector-level distribution of $\text{M}_{\mu\mu}$ after detector-level cuts in \autoref{Table:cutflow_table_Run2} and \autoref{Table:cutflow_table_HLLHC}. The only differences between Run-2 and HL-LHC benchmark are the muon $p_T$ threshold and the muon rapidity coverage at detector-level.}
    \label{fig:mumu_distribution}
\end{figure}

The BSM model file to produce sample events via {\tt MadGraph5\_aMC@NLO} is obtained by using {\tt FeynRules}~\cite{Alloul:2013bka}. Among the simulation work, the parton-level cuts are the same as those in \autoref{sec:parton_lv}. For detector-level analysis, we modify the cut-based analysis to incorporate new resonances. After requiring $80<\text{M}_{\mu\mu\gamma}<100~\text{GeV}$, we require another mass window cut on dimuon invariant mass $\text{M}_{\mu\mu}$ centered at $m_a$. The width of dimuon mass window is $\pm\{ 1.5,2,3,5,\cdots,5 \}~\text{GeV}$ with ALP mass $m_a=\{ 5,15,25,35,\cdots,85 \}~\text{GeV}$. The dimuon mass distribution is also shown in \autoref{fig:mumu_distribution}. Note that from the BSM Feynman diagram, the photon's transverse momentum always orients back-to-back with the dimuon $p_T$. In contrast, within the SM $Z$ boson decay, the photon tends to be emitted collinearly with one of the muons. This distinct difference in event topology offers a handle for discriminating between signal and background. In our cutflow analysis, we observe a strong correlation between the dimuon invariant mass $\text{M}_{\mu\mu}$ and the angular separation $\Delta R(\gamma,\mu)$. As a result, selecting events based on the dimuon resonance effectively incorporates much of the kinematic discrimination. We keep the analysis simple here and leave the space for real experimental data analysis with other multi-variable optimization. Another potential background arises from the process $q\bar{q} \to \gamma + Z/\gamma^*(\to \mu\mu)$, which exhibits kinematics similar to the signal. However, it is significantly suppressed by the requirements on the dimuon invariant mass $\text{M}_{\mu\mu}$, and the angular separation $\Delta R(\mu\mu)$. 
Finally, we choose benchmark points with $m_a>5$~GeV. The lower-mass regime remains accessible, since low-mass dimuon can still be sufficiently boosted to pass triggers, and can be explored with smaller $\Delta R$ cuts together with a dedicated treatment of SM backgrounds, including vector-meson resonances, such as $J/\psi$. 

\begin{figure}[H]
    \centering
    \includegraphics[width=0.49\textwidth]{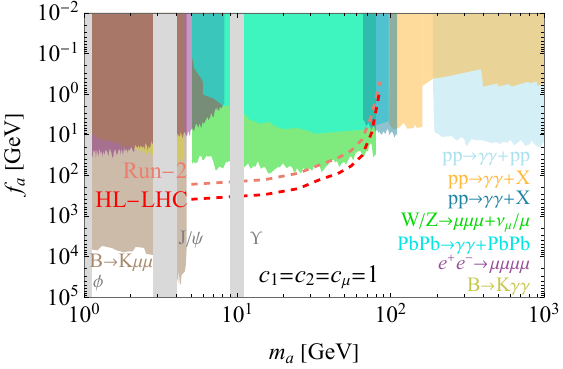}
    \includegraphics[width=0.49\textwidth]{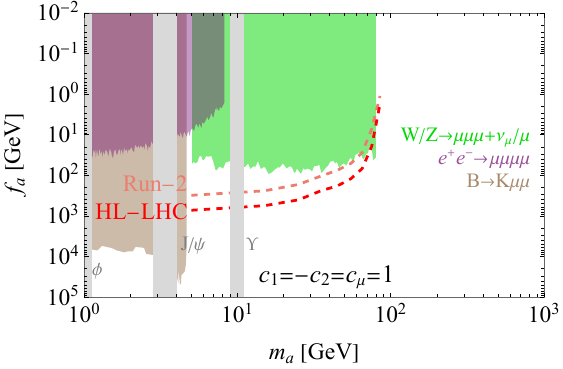}\\
    \includegraphics[width=0.49\textwidth]{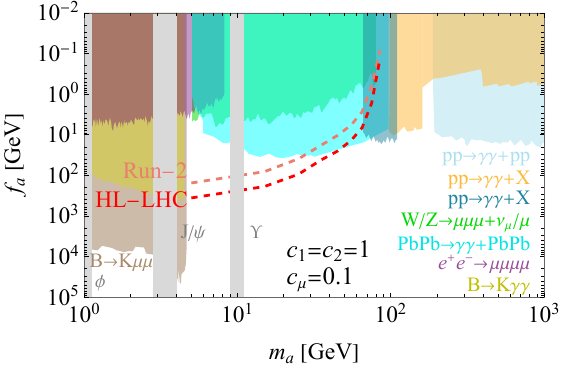}
    \includegraphics[width=0.49\textwidth]{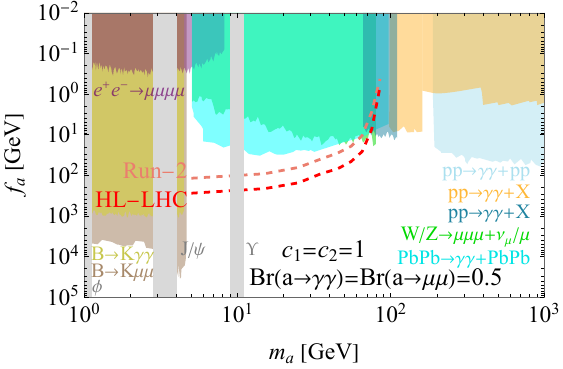}
    \caption{The projected sensitivities of ALP at LHC are shown as red-dashed lines. In the upper-left corner, the purple shaded region is the constraint via $e^+e^-\to 4\mu$ search in Babar~\cite{BaBar:2016sci}. The yellow and brown shaded regions corresponds to $B\to K\gamma\gamma/K\mu\mu$ studies in Babar, Belle and LHCb~\cite{BaBar:2021ich,Belle:2025zbq,Bauer:2021mvw,Izaguirre:2016dfi,LHCb:2016awg}. The green shaded region in the middle is translated from $Z'\to \mu\mu$ searches via $Z\to \mu\mu Z'$ and $W\to\mu\nu_\mu Z'$ at the LHC~\cite{ATLAS:2024uvu,CMS:2018yxg,Fei:2024qtu}. The light-blue shaded region on the right side is an exclusive diphoton resonance search at LHC via photon fusion (requiring a forward proton in the final state)~\cite{ATLAS:2017ayi}. 
    The orange shaded areas also refer to diphoton search at LHC projected by~\cite{Jaeckel:2012yz}. The cyan shaded region comes from $\gamma\gamma\to a\to \gamma\gamma$ studies~\cite{Knapen:2016moh,CMS:2018erd,ATLAS:2020hii} at Pb-Pb collision. In addition, we translate the latest diphoton studies~\cite{ATLAS:2024bjr,CMS:2024yhz} around 95~GeV shown as dark-blue shaded area. The gray shaded regions refer to the $\phi,J/\psi,\Upsilon$ resonances that require additional background study.}
    \label{fig:projection}
\end{figure}

The 95\% C.L. projected sensitivities from this $Z$ radiative decay channel on $f_a$ versus the ALP mass are shown in \autoref{fig:projection} with light-red and red dashed lines for our Run-2 analysis and HL-LHC, respectively. We also show the current exclusions from other existing searches. Different choices of Wilson coefficients are considered to highlight the difference between diphoton and dimuon searches. For example, in the top-right panel, there is no $a F\tilde{F}$ interaction when $c_1=-c_2=1$, yet the ALP can still be produced via radiative $Z$ decay. In the top-left panel, our study exhibits significantly better sensitivity than the diphoton search, simply because the branching ratio $\text{Br}(a\to\mu\mu)$ is much larger than $\text{Br}(a\to\gamma\gamma)$. In the bottom-left panel, the dimuon channel continues to outperform in the low-mass regime, even when $c_\mu$ is an order of magnitude smaller than $c_1,c_2$. Particularly, in the bottom-right panel, we set $c_1=c_2=1$ when calculating the ALP production rate, but fix the ALP decay branching ratios to 50\% dimuon and 50\% diphoton (by adjusting $c_\mu$) to provide a fair comparison of the discovery potentials of the two channels. Finally, we shade several gray bands around $\phi,J/\psi,\Upsilon$ resonances where SM vector-meson backgrounds contribute. A reliable estimate of this background requires a dedicated analysis which is beyond the scope of this work.

\section{Anomalous Dark Force}
\label{sec:dark_U1}
In addition to the ALP scenario, we also consider another type of BSM framework motivated by a gauged dark force that appears to have anomaly at low energy, but is canceled above the electroweak scale~\cite{Dror:2017nsg,Dobrescu:2014fca,Anastasopoulos:2006cz,Dedes:2012me,Arcadi:2017jqd,Ismail:2017ulg,DHoker:1984mif,DHoker:1984izu,Michaels:2020fzj}. As a specific example in mind, one can consider light vector particles coupled to the right-handed muon currents \cite{Batell:2011qq,McKeen:2013dma}. Since leptons are chiral fermions in the SM, introducing a new gauge force that couples to the muon would render the theory gauge anomalous. Consequently, the consistency of the model requires the presence of additional fermions to cancel the resulting gauge anomaly. In the low energy regime, the behavior is determined by this anomalous coupling, incarnated as a dominant contribution from the longitudinal gauge boson, which can also be captured by a Goldstone treatment.
\begin{figure}[ht]
    \centering
    \includegraphics[width=0.24\textwidth]{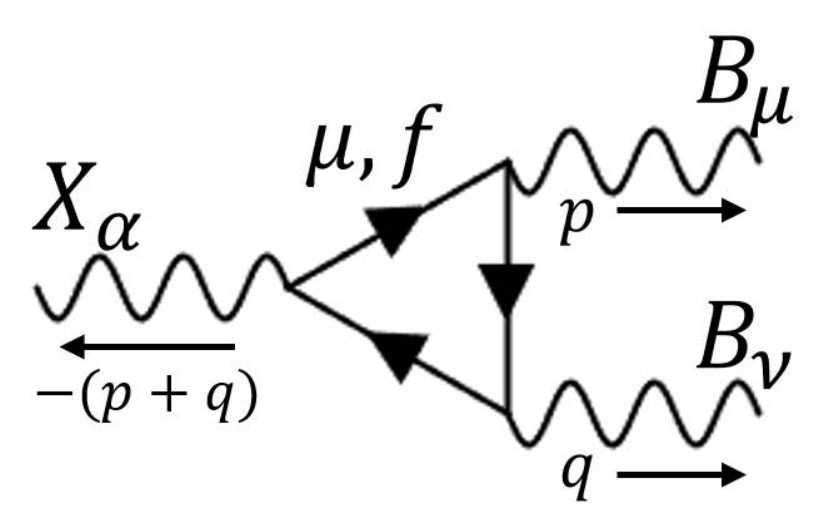}~~
    \includegraphics[width=0.24\textwidth]{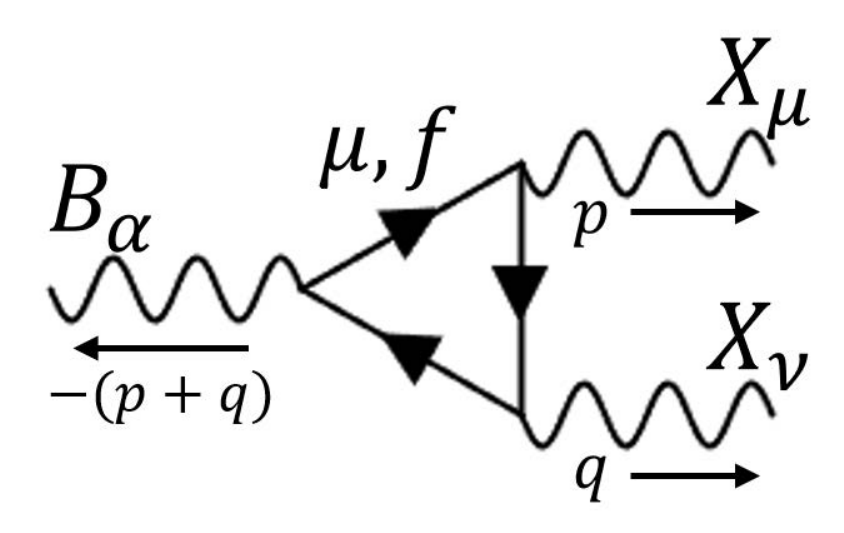}~~
    \includegraphics[width=0.24\textwidth]{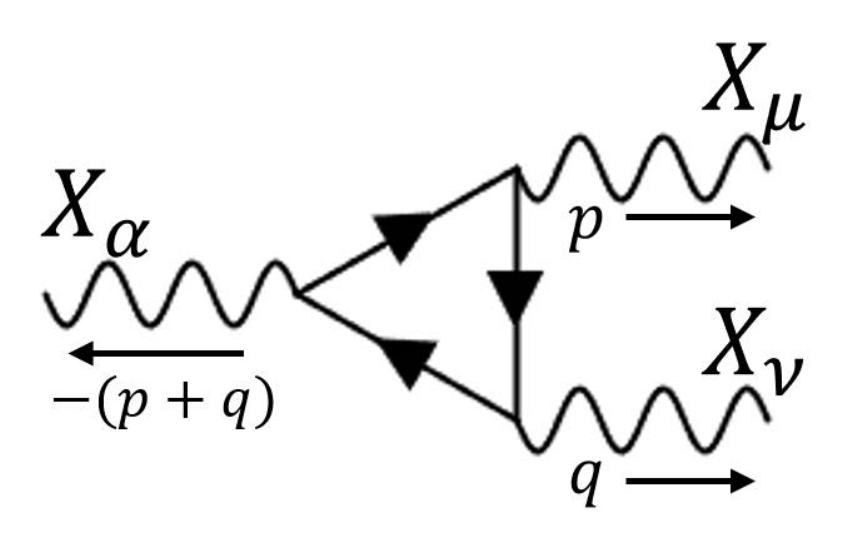}
    \caption{Anomaly diagrams involving $U(1)_X$ and SM $U(1)_Y$ , induced by loops of the new heavy fermion $f$ and the muon.}
    \label{fig:anomaly_diagram}
\end{figure}

In this section, the benchmark model contains a massive dark $U(1)_X$ gauge field and a heavy Dirac fermion, both of which obtain their masses, $m_X$ and $m_f$, through the spontaneous symmetry breaking of a BSM scalar (SM singlet). The gauge boson $X$ couples to the right-handed muon as well as to the heavy fermion $f$. As a result, the $Z$ boson can decay into $\gamma X$ via fermion loop (see the right panel of \autoref{fig:feynman_diagram}), and $X$ could further decay into dimuon. The gauge anomaly arising in the fermion loop (independent of fermion mass) is canceled between the muon and the new fermion $f$. This fermion is assumed to be vector-like with respect to the SM gauge group, and chiral with respect to $X$. Therefore, the same mechanism that  renders $f$ heavy will also provide the mass to $X$ boson. New fermion $f$ will have to be heavy, or at least above the electroweak scale, as it will carry the electric charge.  It is well appreciated in the literature that the $f$ loop does not fully decouple, leading to a non-vanishing amplitude for $Z \to \gamma X$. The relevant Lagrangian can be extracted from~\cite{Dror:2017nsg}: 
\begin{align}
    \mathcal{L}_{X,f}\supset Q_X^\text{SM}g_X X_\alpha(\bar{\mu}\gamma^\alpha P_R \mu ) + Q_X^f g_X  X_\alpha\bar{f}\gamma^\alpha(c_v+c_A\gamma_5) f + Y_f g'\bar{f}\slashed{B}f 
\end{align}
where $P_R=\frac{1+\gamma_5}{2}$, $B^\alpha$ is the SM $U(1)_Y$ field and $Q_X^\text{SM},Q_X^f$ are $U(1)_X$ the charges of $\mu_R, f$. Because the new gauge interaction couples only to the right-handed muon\footnote{The Yukawa interaction of muon needs to be replaced by a high-dimensional operator. We discuss the UV completion in \autoref{sec:uv_complete}.}, there is no gauge anomaly associated with the SM $SU(2)_L$ field. Moreover, since the $U(1)_Y$ gauge boson couples vectorially to $f$, the new fermion does not induce a $U(1)_Y^3$ anomaly. However, three possible gauge anomalies can arise: $U(1)_X U(1)_Y^2$, $U(1)_X^2 U(1)_Y$, and $U(1)_X^3$, as illustrated in \autoref{fig:anomaly_diagram}. The $U(1)_X^3$ anomaly can always be canceled by introducing additional degrees of freedom that are charged only under $U(1)_X$. Therefore, the anomalies relevant for our setup are $U(1)_X U(1)_Y^2$ and $U(1)_X^2 U(1)_Y$:
\begin{align}
    -(p+q)_\alpha \mathcal{M}^{\alpha\mu\nu}_A&=\frac{Y_{\mu_R}^2Q_X^\text{SM}+2Y_f^2 Q_X^f c_A }{12\pi^2}g_X{g'}^2p_\sigma q_\rho \epsilon^{\sigma\rho\mu\nu}\, ,\\
     -(p+q)_\alpha \mathcal{M}^{\alpha\mu\nu}_A&=\frac{Y_{\mu_R}(Q_X^\text{SM})^2+Y_f (Q_X^f)^2(c_v+c_A)^2-Y_f (Q_X^f)^2(c_v-c_A)^2}{12\pi^2}g_X^2{g'}p_\sigma q_\rho \epsilon^{\sigma\rho\mu\nu}\, ,
\end{align}
where $\mathcal{M}_A$ denotes the gauge-anomalous contribution, which is independent of the fermion mass, and $Y_{\mu_R}=-1$ is the hypercharge of the right-handed muon. To ensure anomaly cancellation, the charge assignments must be chosen appropriately. For example, $Y_f=2c_A=2c_v=-Q_X^f=Q_X^\text{SM}=1$ would make the model anomaly-free. As discussed below, although there are different solutions to assign charges, the amplitude of $Z \to \gamma X$ is unchanged once we choose to normalize $Q_X^\text{SM}=1$. 

The divergence of the $XBB$ triangle diagram is 
\begin{align}
    -(p+q)_\alpha \mathcal{M}^{\alpha\mu\nu}&=\frac{Y_{\mu_R}^2Q_X^\text{SM}+2Y_f^2 Q_X^f c_A }{12\pi^2}g_X{g'}^2p_\sigma q_\rho \epsilon^{\sigma\rho\mu\nu}+\frac{2Y_f^2 Q_X^f c_A}{4\pi^2}g_X{g'}^2p_\sigma q_\rho \epsilon^{\sigma\rho\mu\nu}\left[2m_f^2 I_{00}(m_f,p,q)\right]
\end{align}
where the second term represents the mass-dependent contribution, with
\begin{align}
    2m_f^2 I_{00}(m_f,p,q)=\int_0^1 dx \int_0^{1-x} dy \frac{2m_f^2}{y(1-y)p^2+x(1-x)q^2+2xy~p\cdot q - m_f^2}\sim 
    \begin{cases}
        -1\quad, ~m_f^2\gg p^2,q^2,p\cdot q \\
        ~0~ \quad, ~m_f^2\ll p^2,q^2,p\cdot q
    \end{cases}.
\end{align}
The first gauge-anomalous contribution must vanish, which implies the condition $2Y_f^2 Q_X^f c_A=-Y_{\mu_R}^2Q_X^\text{SM}$. Consequently, the final result of the $XBB$ diagram (for heavy $f$) is
\begin{align}
    -(p+q)_\alpha \mathcal{M}^{\alpha\mu\nu}=\frac{Y_{\mu_R}^2Q_X^\text{SM}}{4\pi^2}g_X{g'}^2p_\sigma q_\rho \epsilon^{\sigma\rho\mu\nu}=\frac{1}{4\pi^2}g_X{g'}^2p_\sigma q_\rho \epsilon^{\sigma\rho\mu\nu}\, . \label{eq:divergence}
\end{align}
The three-point amplitude of $X_L BB$ can be written as $\frac{(p+q)_\alpha}{m_X}\mathcal{M}^{\alpha\mu\nu}\epsilon_\mu(p) \epsilon_\nu(q)$, since $\epsilon_L\approx\frac{k^\mu}{m}$ in high energy limit. Therefore, the amplitude of $Z\to \gamma X$ is fully determined by \autoref{eq:divergence}, even though there are multiple solutions to assign charges within this model. The corresponding decay width of $Z\to \gamma X_L$ is given by
\begin{align}
    \Gamma(Z\to \gamma X_L)\approx \frac{g_X^2{g'}^2g^2}{1536\pi^5}\frac{m_Z^3}{m_X^2}(1-m_X^2/m_Z^2)^3~.
\end{align}

In an alternative (and equivalent) language \cite{Dror:2017nsg}, $Z$ decays to a photon and the Goldstone boson associated with the Higgs mechanism for $X$ vector. The vector nature of $X$-particle is important for its decay: {\em i.e.} in contrast to a pure ALP case, the decay to photons is forbidden, and is expected to occur to muons in this model. 

\begin{figure}[ht]
    \centering
    \includegraphics[width=0.7\textwidth]{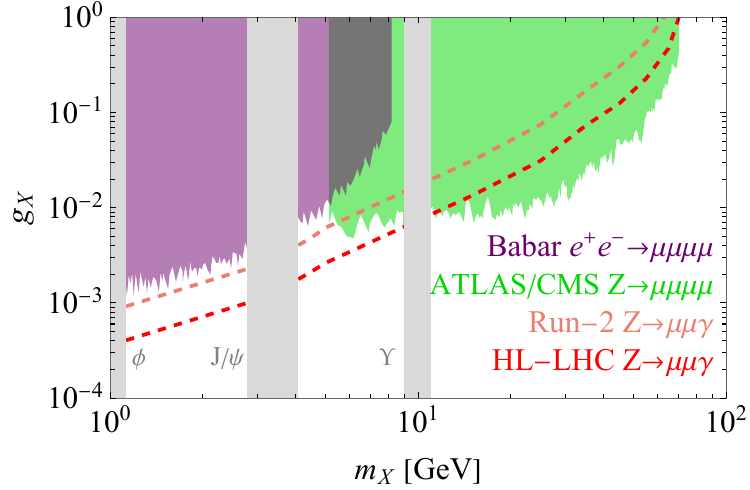}
    \caption{The projected sensitivities of the dark gauge $U(1)_X$ field are shown as red-dashed lines. The purple shaded region is the constraint via $e^+e^-\to 4\mu$ search in Babar~\cite{BaBar:2016sci}. The green shaded region is translated from $Z'\to \mu\mu$ searches via $Z\to \mu\mu Z'$~\cite{ATLAS:2024uvu,CMS:2018yxg}. The gray shaded regions refer to the $\phi,J/\psi,\Upsilon$ resonances that require additional background study.}
    \label{fig:projection_U1}
\end{figure}

Consequently, we focus on the longitudinal $X$ production channel, which benefits from the $m_Z^2/m_X^2$ enhancement.
As shown by the red dashed lines in \autoref{fig:projection_U1}, the projected sensitivities improve as $m_X$ decreases. In general, the four-muon final state provides higher sensitivity than the $Z\to\gamma X_L$ channel, which suffers from loop suppression.\footnote{\textcolor{red}{The decay $Z\to XX\to 4\mu$ could also provide an interesting complementary discovery channel with an effective $ZXX$ interaction.}} However, the high luminosity of the LHC allows it to achieve better performance in the low-mass region compared to the PEP-II $e^+e^-$ collider. For $m_X>5~\text{GeV}$, the sensitivity is directly translated from our ALP study. In the low-mass region $1~\text{GeV}< m_X \leq 5~\text{GeV}$, we gradually enlarge the mass window from $\pm 0.1~\text{GeV}$ to $\pm 0.5~\text{GeV}$, for which both ATLAS and CMS have sufficient resolution in the dimuon invariant mass~\cite{CMS:2017dju,ATLAS:2022rej}. At such low masses, complicated background from hadron decays become significant, and a data-driven approach would provide a robust strategy to search for a dimuon resonance. 
For simplicity, we just assume the background $\text{M}_{\mu\mu}$ distribution below $5~$GeV, $\frac{1}{\sigma}\frac{d\sigma}{d\text{M}_{\mu\mu}}$, remains approximately constant at its value near $\text{M}_{\mu\mu}=5~\text{GeV}$. In addition, for $m_X<5~\text{GeV}$, the detection efficiency is assumed to be the same as that at $m_X=5~\text{GeV}$.

\section{Summary}
\label{sec:conclusions}
In this work, we investigated the radiative decay of the $Z$ boson, $Z \to \mu^+\mu^-\gamma$, at the LHC. This decay channel, while present in the Standard Model, had only limited experimental scrutiny at LEP. Leveraging the large $Z$ boson samples produced at the LHC, we provided updated projections for the measurement of its fiducial branching ratio using both Run-2 and HL-LHC detector benchmarks. Our results demonstrate that the $Z \to \mu^+\mu^-\gamma$ channel can be measured with high statistical precision, at the level of $\mathcal{O}(0.1\%)$, making it a valuable target for precision tests of the electroweak sector. Additionally, we extract the fiducial branching ratio from a Run-1 study, $\text{Br}^\text{fid}=(3.34\pm0.016)\times 10^{-4}$. A separate analysis of the systematic uncertainty by the experimental collaborations is needed for extracting the result with the full experimental error.

The discovery potentials for BSM scenarios were also explored in the context of ALPs and an anomalous $U(1)_X$ gauge force that couples to the muon. The $Z \to a/X_L + \gamma \to \mu^+\mu^-\gamma$ process provides a distinctive signature, where the dimuon invariant mass exhibits a narrow resonance over the SM background. We performed a dedicated simulation and cut-based analysis, projecting the sensitivity of probing these new particles in the mass range of 1$-$90~GeV. Our results indicate that the radiative $Z$ decay offers a competitive and complementary probe to existing diphoton studies, particularly for the ALP searches. For the dark $U(1)_X$ gauge field, the LHC performs better than the existing constraint in the low-mass region, probing the dark gauge coupling $g_X$ down to $O(10^{-3})$. 
Our results highlight the potential of rare electroweak boson decays as sensitive probes of new physics, motivating further phenomenological and experimental studies of both $Z$ and $W$ decay modes at the LHC.

\acknowledgments

We thank Matthew Herndon for useful discussions. M.P. is thankful to Drs. M. Baker and A. Thamm for initial discussions of $Z$ decay sensitivity to ALPs.  Y.F. is supported by the National Natural Science Foundation of China under Grant No.~12035008 and No.~12375091 at Fudan University. P.L., Z.L. and M.P. are supported by the Department of Energy under Grant No.~DE-SC0011842 at the University of Minnesota. P.L. is partly supported by a Doctoral Dissertation Fellowship at the University of Minnesota. Z.L. is supported in part by a Sloan Research Fellowship from the Alfred P. Sloan Foundation at the University of Minnesota. 
The data and core codes for the research are publicly available at \href{https://github.com/ivanfei-1/Ztomumugamma_LHC}{Github~\faGithub}.

\appendix
\section{UV completion of anomalous muonic dark force}
\label{sec:uv_complete}
In this appendix, we provide a UV-complete model for the anomalous $U(1)_X$ benchmarks. Since the right-handed muon is the only SM Weyl fermion charged under $U(1)_X$, the SM Yukawa term $\bar{L}_\mu H \mu_R$ is forbidden. To generate the muon mass, we introduce the following BSM Yukawa interactions:
\begin{align}
    \mathcal{L}^\text{BSM}_\text{Yukawa}\supset y_1 \bar{L}_\mu \tilde{H} N_R + y_2\Phi \bar{N}_L N_R + y_3\phi \bar{N}_L\mu_R + \text{h.c.}\, ,
\end{align}
where $\phi$ is a real scalar carrying $U(1)_{Y/X}$ charges opposite to those of $\mu_R$, while $\Phi$ and the Dirac fermion pair $N_{L,R}$ are singlets under the SM gauge group $SU(3)_c \times SU(2)_L \times U(1)_Y$ and $U(1)_X$.

\begin{figure}
    \centering
    \includegraphics[width=0.3\linewidth]{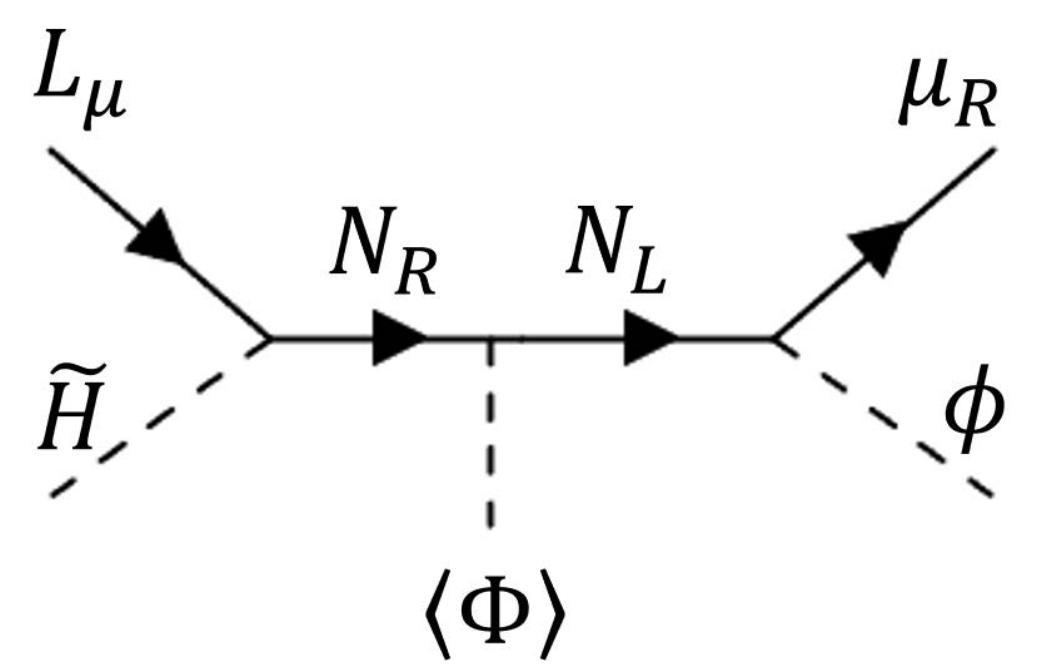}
    \caption{The Feynman diagram of the dimension-5 operator (\autoref{eq:EFT_op}) generated by integrating out the heavy Dirac fermion $N$.}
    \label{fig:uv_diagram}
\end{figure}

Integrating out the heavy Dirac fermion $N$ generates an effective dimension-5 operator,
\begin{align}
\mathcal{O}_\text{EFT}
= y_1 y_2 y_3 \frac{\phi \langle \Phi \rangle\bar{L}_\mu \tilde{H} \mu_R  }{m_N^2}
=y_1 y_2 y_3 \frac{\phi \langle \Phi \rangle\bar{L}_\mu \tilde{H} \mu_R  }{y_2^2 \langle \Phi \rangle^2}= \frac{y_1 y_3}{y_2} \frac{\phi}{\langle \Phi \rangle}
\bar{L}_\mu \tilde{H} \mu_R\, , \label{eq:EFT_op}
\end{align}
which effectively reproduces the SM Yukawa coupling once $\phi$ acquires a vacuum expectation value, thereby spontaneously breaking the $U(1)_X$ symmetry and giving mass to the $X$ gauge boson (and the heavy fermion $f$). The corresponding Feynman diagram is shown in \autoref{fig:uv_diagram}.

\bibliographystyle{utphys}
\bibliography{references}

\end{document}